\documentclass[conference]{IEEEtran}
\IEEEoverridecommandlockouts
\usepackage{cite}
\usepackage{amsthm, amsmath,amssymb,amsfonts}
\usepackage{algorithmic}
\usepackage{graphicx}
\usepackage{mathrsfs}
\usepackage{stmaryrd}
\usepackage{textcomp}
\usepackage{xcolor}
\usepackage{url}
\usepackage{tikz}
\usepackage{scalerel}
\usepackage[english]{babel}
\usetikzlibrary{matrix, positioning, arrows.meta}
\usepackage{adjustbox}
\usepackage{commath}
\def\BibTeX{{\rm B\kern-.05em{\sc i\kern-.025em b}\kern-.08em
    T\kern-.1667em\lower.7ex\hbox{E}\kern-.125emX}}

\newtheorem{theorem}{Theorem}[section]

\newtheorem{definition}[theorem]{Definition}
\newtheorem{remark}[theorem]{Remark}
\newtheorem*{example}{Example}

\newcommand{\indep}{\perp \!\!\! \perp}

\newcommand{\sbrackets}[1]{\left[#1\right]}

\begin{document}

\title{Causal Vector-valued Witsenhausen Counterexamples with Feedback\\
\thanks{This work was supported by Swedish Research Council (VR) under grant 2020-03884.}
}

\author{\IEEEauthorblockN{Mengyuan Zhao}
\IEEEauthorblockA{Division of Information Science\\and Engineering\\
KTH Royal Institute of Technology\\
100 44 Stockholm, Sweden\\
Email: mzhao@kth.se}
\and
\IEEEauthorblockN{Maël Le Treust}
\IEEEauthorblockA{Univ. Rennes, CNRS, Inria\\ IRISA UMR 6074\\
F-35000 Rennes, France\\
Email: mael.le-treust@cnrs.fr}
\and
\IEEEauthorblockN{Tobias J. Oechtering}
\IEEEauthorblockA{Division of Information Science\\and Engineering\\
KTH Royal Institute of Technology\\
100 44 Stockholm, Sweden\\
Email: oech@kth.se}}
\maketitle

\begin{abstract}
We study the continuous vector-valued Witsenhausen counterexample through the lens of empirical coordination coding. We characterize the region of achievable pairs of costs in three scenarios: (i) causal encoding and causal decoding, (ii) causal encoding and causal decoding with channel feedback, and (iii) causal encoding and noncausal decoding with channel feedback. In these vector-valued versions of the problem, the optimal coding schemes must rely on a time-sharing strategy, since the region of achievable pairs of costs might not be convex in the scalar version of the problem. We examine the role of the channel feedback when the encoder is causal and the decoder is either causal or non-causal, and we show that feedback improves the performance, only when the decoder is non-causal.
\end{abstract}


\section{Introduction}
In 1968, Witsenhausen introduced his famous counterexample demonstrating the suboptimality of affine strategies in the Linear Quadratic Gaussian (LQG) settings featuring non-classical information patterns \cite{witsenhausen1968}. Since then, it has become a cornerstone in the field of distributed decision-making \cite{bansal1986stochastic, silva2010control, yuksel2013stochastic, gupta2015existence} and information-theoretic control \cite{martins2005fundamental,
freudenberg2008feedback, 
derpich2012improved, AgrawalDLL15, Akyol2017information, charalambous2017hierarchical, wiese2018secure, 
Stavrou2022sequential}.


Due to the vector-valued formulation of the problem \cite{Grover2010Witsenhausen}, many information-theoretic approaches have been adopted for analyzing this open problem \cite{elgamal2011nit, kim2008state,sumszyk2009information, choudhuri2013causal}. The concept of empirical coordination, introduced in \cite{cuff2010coordination, cuff2011hybrid, cuff2011coordination}, plays an important role in establishing cooperative behavior among all agents in the network, providing single-letter solutions for problems such as characterizing the optimal cost, capacity region, and utility functions \cite{Treust2017joint , larrousse2015coordinationISIT, larrousse2018coordination}. 


Feedback, as an intrinsic component of communication systems, offers the potential to enhance performance by enabling decision makers (DMs) to refine their actions based on previous outcomes. In many multi-terminal setups, feedback has proven beneficial for increasing the capacity region and assisting communication of the multiple-access channel \cite{gaarder1975capacity,ozarow1984capacity} as well as broadcast channel \cite{ozarow1984achievable, dueck1979capacity}. Additionally, when considering  the empirical coordination coding problem in a point-to-point scenario, \cite{Letreust2015empirical, Letreust2021state} showed that channel feedback could enable the DMs to directly coordinate their outputs with the system state, by simplifying the information constraint and by reducing the number of auxiliary random variables, and therefore, enlarge the set of distributions. However, whether feedback facilitates coordination in the vector-valued Witsenhausen settings with causal controllers \cite{Treust2024power, zhao2024coordination} still remains an open question. Recently, it has been claimed in \cite{zhao2024CDC}, that when constraint to Gaussian case, feedback does not contribute to performance improvements when the first DM is causal and the second DM is noncausal. In this article, we complete this study and we provide the necessary proof to the claimed information constraint.

In this paper, we aim to deepen the understanding of the problem by incorporating channel feedback for a comparative study, focusing on three novel setups: (i) causal encoding and causal decoding, (ii) causal encoding and causal decoding with channel feedback, and (iii) causal encoding and noncausal decoding with channel feedback. Our analysis shows that time-sharing is crucial in the first setup, in order to randomize between operating points and convexify the region of achievable pairs of costs. The minimum Gaussian cost identified in \cite{zhao2024CDC} can also be achieved by time-sharing between two affine policies. Time-sharing corresponds to randomized decision-making process for the original one-shot Witsenhausen counterexample which enables operating points to be achieved by randomization. The characterization of the region of achievable pairs of costs for the second model is derived using the genie-aided argument by focusing on its outer-bound. Comparing the results of the second and third models, we conclude that feedback can enlarge the region of achievable pairs of costs  when the decoder is non-causal but not when it is causal.

\section{System Model}\label{sec: system model}
\begin{figure}[!t]
  \centering


\begin{tikzpicture}[scale=0.9, every node/.style={scale=0.9}]
    \draw (2,0) rectangle (3,1);
    \draw (6.8,0) rectangle (7.8,1);

    \draw (4.2,0.5) circle (0.2) node {$+$};
    \draw (5.6,0.5) circle (0.2) node {$+$};

    \filldraw (1,1.5) circle (2pt) node[above] {$X_{0,t}\sim \mathcal{N}(0,Q)$};
    \filldraw (5.6,1.5) circle (2pt) node[above] {$Z_{1,t}\sim \mathcal{N}(0,N)$};

    \draw[->] (1,1.5) -- (1,0.5) -- (2,0.5);
    \draw[->] (1,1.5) -- (4.2,1.5) -- (4.2,0.7);
    \draw[->] (3,0.5) -- (4,0.5);
    \draw[->] (4.4,0.5) -- (5.4,0.5);
    \draw[->] (4.9,0.5) -- (4.9,-0.5) -- (8.8,-0.5);
    \draw[->] (5.6,1.5) -- (5.6,0.7);
    \draw[->] (5.8,0.5) -- (6.8,0.5);
    \draw[->] (7.8,0.5) -- (8.8,0.5);

    \node at (1.5,0.8) {$X_0^t$};
    \node at (3.5,0.8) {$U_{1,t}$};
    \node at (4.9,0.8) {$X_{1,t}$};
    \node at (6.3,0.8) {$Y_{1}^t$};
    \node at (8.3,0.8) {$U_{2,t}$};
    \node at (8.3,-0.2) {$X_{1,t}$};
    \node at (2.5,0.5) {$C_1$};
    \node at (7.3,0.5) {$C_2$};
\end{tikzpicture}
\caption{Vector-valued Witsenhausen counterexample with causal encoder and causal decoder.}
\label{fig:model}

  \label{fig: c-c wits prob}
\end{figure}
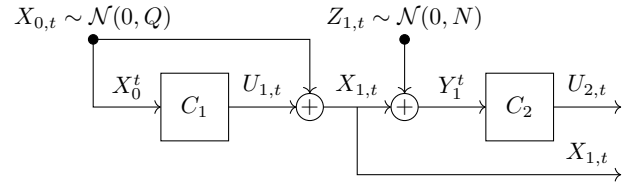

The vector-valued Witsenhausen counterexample setup consists of source states and channel noises that are drawn independently according to the i.i.d. Gaussian distributions $X_0^n\sim\mathcal{N}(0, Q\mathbb I)$ and $Z_1^n\sim\mathcal{N}(0,N\mathbb I)$, for some block length $n>0$, and $Q,N>0$. We denote by $X_1$ the memoryless interim state and $Y_1$ and output of the memoryless additive channel, generated by
\begin{align}
    &X_1 = X_0 + U_1  &\text{with }X_0\sim\mathcal{N}(0,Q),\label{X1 generation}\\
    &Y_1 = X_1 + Z_1 = X_0 + U_1 + Z_1 &\text{with }Z_1\sim\mathcal{N}(0,N).\label{Y1 generation}
\end{align}
We denote by $\mathcal{P}_{X_0} = \mathcal{N}(0,Q)$ the generative Gaussian probability distribution of the state, and by $\mathcal{P}_{X_1,Y_1|X_0,U_1}$ the channel probability distribution according to \eqref{X1 generation} and \eqref{Y1 generation}.

We now present the three models and their corresponding results. The proofs of these results are given in Appendix.
\subsection{Causal encoding and Causal decoding}

Let's first consider the model with two causal DMs, as illustrated in Fig. \ref{fig: c-c wits prob}.

\begin{definition}
    For $n\in\mathbb{N}$, a ``control design'' with causal encoder and causal decoder is a tuple of stochastic functions $c = (\{ f^{(t)}_{U_{1,t}|X_0^t}\}_{t=1}^n, \{ g^{(t)}_{U_{2,t}|Y_1^t}\}_{t=1}^n)$ defined by
    \begin{equation}
        f^{(t)}_{U_{1,t}|X_0^t}: \mathcal{X}_0^t \longrightarrow \mathcal{U}_{1,t},\quad g^{(t)}_{U_{2,t}|Y_1^t}: \mathcal{Y}_1^t\longrightarrow \mathcal{U}_{2,t},
    \end{equation}
    which induces a distribution over sequences of symbols:
    \begin{equation}
         \prod_{t=1}^n \mathcal{P}_{X_{0,t}} \prod_{t=1}^n f^{(t)}_{U_{1,t}|X_0^t}\prod_{t=1}^n \mathcal{P}_{X_{1,t},Y_{1,t}|X_{0,t},U_{1,t}} \prod_{t=1}^n g^{(t)}_{U_{2,t}|Y_1^t}. \label{eq: distribution of sequences}
    \end{equation}
We denote by $\mathcal{C}_{ed}(n)$ the set of control designs with causal encoder and causal decoder.
\end{definition} 

\begin{definition}\label{def: achievable cost}
    We define the two long-run cost functions $c_P(u_1^n) = \frac{1}{n}\sum_{t=1}^n (u_{1,t})^2$ and $c_S(x_1^n, u_2^n) = \frac{1}{n}\sum_{t=1}^n(x_{1,t}-u_{2,t})^2$. 
    The pair of costs $(P,S)\in\mathbb{R}^2$ is said to be achievable if for all $\varepsilon>0$, there exists $\Bar{n}\in\mathbb N$ such that for all $n\geq \Bar{n}$, there exists a control design $c\in \mathcal{C}_{ed}(n)$ such that 
    \begin{equation}
        \mathbb E\Big[\big|P - c_P(U_1^n)\big| + \big|S - c_S(X_1^n, U_2^n)\big|\Big] \leq \varepsilon.\label{eq: achievable def}
    \end{equation}
    We denote by $\mathcal{R}_{ed}$ the set of achievable pairs of costs for control designs in $\mathcal{C}_{ed}$.
\end{definition}

Next, we characterize the costs region $\mathcal{R}_{ed}$.

\begin{theorem}\label{theorem: c-c wits main}
    The pair of Witsenhausen costs $(P,S)$ is achievable if and only if there exists a joint distribution over the random variables $(X_0, T, U_1, X_1, Y_1, U_2)$ that decomposes according to
    \begin{equation}
\mathcal{P}_{X_0}\mathcal{P}_{T}\mathcal{P}_{U_1|X_0,T}\mathcal{P}_{X_1, Y_1|X_0, U_1}\mathcal{P}_{U_2|T, Y_1},\label{c-c prob result}
    \end{equation}
    such that
\begin{align}
 P = \mathbb{E}\sbrackets{U_1^2}, \quad \quad S = \mathbb{E}\sbrackets{(X_1 - U_2)^2},\label{eq: c-c cost result}
\end{align}
    where $\mathcal{P}_{X_0}$ and $\mathcal{P}_{X_1, Y_1|X_0, U_1}$ are the given Gaussian distributions, and $T$ is the time-sharing auxiliary random variable with cardinality bound $|\mathcal{T}|\leq 2$.
\end{theorem}

\begin{remark}
    The probability distribution \eqref{c-c prob result} satisfies
    \begin{align}
        \left\{  
        \begin{aligned}
            &X_0\text{  is independent of  }T,\\
            &(X_1, Y_1)-\!\!\!\!\minuso\!\!\!\!- (X_0, U_1)    -\!\!\!\!\minuso\!\!\!\!- T,\\
            & U_2 -\!\!\!\!\minuso\!\!\!\!- (T, Y_1) -\!\!\!\!\minuso\!\!\!\!- (X_0, U_1, X_1).
        \end{aligned}
        \right.
        \label{markov result}
    \end{align}
 The first property comes from the fact that the time-sharing random variable is independent of the source. The second Markov chain is due to the memoryless property of the channel. The last Markov chain comes from the causal decoding and the symbol-wise reconstruction.
\end{remark}

\begin{remark}
    The cost region 
    \begin{align}
        \mathcal{R}_{ed} &= \{(P,S):P = \mathbb E[U_1^2], S = \mathbb E[(X_1 - U_2)^2],\nonumber\\
        &\quad\quad\text{ for }\mathcal{P} \text{ of the form of \eqref{c-c prob result}}\}
    \end{align}
    characterized in Theorem \ref{theorem: c-c wits main} is convex. This is because time-sharing synthesizes DMs to agree on the operating points. Therefore, we can construct a control scheme that achieves any point on the cord between two pairs of Witsenhausen costs. 
    
    \end{remark}

 Let's look at the follwing example delving into the effect of convexification using time-sharing:
    \begin{example}
        We consider a binary source $X_0$ with $\mathbb{P}(X_0 = 0)=\mathbb{P}(X_0 = 1)=\frac{1}{2}$, binary symmetric stochastic encoder $\sim\text{Bern}(\alpha)$ and decoder $\sim\text{Bern}(\beta)$ and a perfect channel with the set of two symbols $\mathcal{X}_0 = \mathcal{U}_1 = \mathcal{U}_2 = \{0,1\} $, as represented in Fig. \ref{fig: binary example1}.
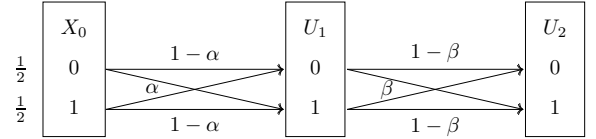
\begin{figure}[ht]
  \centering


\scalebox{0.80}{\begin{tikzpicture}[scale=0.50]
  \tikzset{matrixstyle/.style={matrix of math nodes, outer sep=0pt, text height=1.5ex, text depth=0.5ex, minimum width=2em, draw, very thin, nodes={inner sep=5pt}}}

  \matrix[matrixstyle] (X0) {X_0 \\ 0 \\ 1 \\};
  \matrix[matrixstyle, right=3cm of X0] (U1) {U_1 \\ 0 \\ 1 \\};
  \matrix[matrixstyle, right=3cm of U1] (U2) {U_2 \\ 0 \\ 1 \\};

  \draw[->] ([xshift=1em]X0-2-1.east) -- ([xshift=-1em]U1-2-1.west) node[midway, above] {$1-\alpha$};
  \draw[->] ([xshift=1em]X0-3-1.east) -- ([xshift=-1em]U1-3-1.west) node[midway, below] {$1-\alpha$};
  \draw[->] ([xshift=1em]X0-2-1.east) -- ([xshift=-1em]U1-3-1.west) node[midway] {};
  \draw[->] ([xshift=1em]X0-3-1.east) -- ([xshift=-1em]U1-2-1.west) node[midway] {};

  \draw[->] ([xshift=1em]U1-2-1.east) -- ([xshift=-1em]U2-2-1.west) node[midway, above] {$1-\beta$};
  \draw[->] ([xshift=1em]U1-3-1.east) -- ([xshift=-1em]U2-3-1.west) node[midway, below] {$1-\beta$};
  \draw[->] ([xshift=1em]U1-2-1.east) -- ([xshift=-1em]U2-3-1.west) node[midway] {};
  \draw[->] ([xshift=1em]U1-3-1.east) -- ([xshift=-1em]U2-2-1.west) node[midway] {};

  \node[left=0.3cm of X0-2-1] (p) {$ \frac{1}{2}$};
  \node[left=0.3cm of X0-3-1] (oneMinusP) {$\frac{1}{2}$};
   \node at (2.6,-0.65) {$\alpha$};
    \node at (10.4,-0.67) {$\beta$};
    \end{tikzpicture}}

   \caption{Binary information source and binary symmetric stochastic encoder and decoder.}
  \label{fig: binary example1}
\end{figure}

The joint distribution of this system is presented in Fig. \ref{fig: binary example joint distr}. The goal of the two DMs is to design their strategies through pairs $(\alpha,\beta)$ to achieve a desired empirical distribution. 

\begin{figure}[ht]
  \centering


\begin{tikzpicture}[text depth=0.2ex, text height=2ex, scale=0.4, every node/.style={scale=0.6}]
 \tikzset{matrixstyle/.style={
    matrix of math nodes, nodes in empty cells, 
    outer sep=0pt, 
    text height=2ex, text depth=1ex,  
    minimum width=4em, minimum height=3em,   
    nodes={draw, inner sep=3pt, outer sep=0pt, text width=6.8em, align=center, anchor=center},
    column sep=-\pgflinewidth, row sep=-\pgflinewidth
  }}

  \matrix[matrixstyle] (m1) {
      \frac{1}{2}(1-\alpha)(1-\beta) & \frac{1}{2}(1-\alpha)\beta \\
      \frac{1}{2}\alpha\beta & \frac{1}{2}\alpha(1-\beta) \\
  };
  \node[left=1mm of m1-1-1.west, anchor=east] {$U_1 = 0$};
  \node[left=1mm of m1-2-1.west, anchor=east] {$U_1 = 1$};
  \node[above=2mm of m1-1-1.north, anchor=south] {$U_2 = 0$};
  \node[above=2mm of m1-1-2.north, anchor=south] {$U_2 = 1$};
  \node[below=2mm of m1.south, anchor=north] {$X_0 = 0$};

  \matrix[matrixstyle, right=1cm of m1] (m2) {
      \frac{1}{2}\alpha(1-\beta) & \frac{1}{2}\alpha\beta \\
      \frac{1}{2}(1-\alpha)\beta & \frac{1}{2}(1-\alpha)(1-\beta) \\
  };
  \node[left=1mm of m2-1-1.west, anchor=east] {$U_1 = 0$};
  \node[left=1mm of m2-2-1.west, anchor=east] {$U_1 = 1$};
  \node[above=2mm of m2-1-1.north, anchor=south] {$U_2 = 0$};
  \node[above=2mm of m2-1-2.north, anchor=south] {$U_2 = 1$};
  \node[below=2mm of m2.south, anchor=north] {$X_0 = 1$};

\end{tikzpicture}

%
   \caption{The joint distribution induced by a binary system depicted in Fig. \ref{fig: binary example1}}
  \label{fig: binary example joint distr}
\end{figure}

In this example, we initially don't apply the time-sharing technique. Fig. \ref{fig: binary example, 0,1} shows an empirical distribution (third row) that is generated by directly combining the distributions given by $(\alpha = 0, \beta = 0)$ (first row) and $(\alpha = 1, \beta = 1)$ (second row). However, one can easily show that this combined empirical distribution can not be achieved by any single choice of pair $(\alpha, \beta)\in [0,1]^2$, thus it is achievable only through time-sharing.

\begin{figure}[ht]
\vspace{0.3cm}
  \centering


\begin{tikzpicture}[text depth=0.2ex, text height=2ex, scale=0.8, every node/.style={scale=0.8}]
 \tikzset{matrixstyle/.style={
    matrix of math nodes, nodes in empty cells, 
    outer sep=0pt, 
    text height=2ex, text depth=1ex,  
    minimum width=1em, minimum height=1em,   
    nodes={draw, inner sep=3pt, outer sep=0pt, text width=3em, align=center, anchor=center},
    column sep=-\pgflinewidth, row sep=-\pgflinewidth
  }}

  \matrix[matrixstyle] (m1) {
      \frac{1}{2} &0 \\
       0 & 0 \\
  };
  \node[left=1mm of m1-1-1.west, anchor=east] {$U_1 = 0$};
  \node[left=1mm of m1-2-1.west, anchor=east] {$U_1 = 1$};
  \node[above=1mm of m1-1-1.north, anchor=south] {$U_2 = 0$};
  \node[above=1mm of m1-1-2.north, anchor=south] {$U_2 = 1$};
  \node[below=0.8mm of m1.south, anchor=north] {$X_0 = 0$};

  \matrix[matrixstyle, right=1cm of m1] (m2) {
      0 &0 \\
       0 & \frac{1}{2} \\
  };
  \node[left=1mm of m2-1-1.west, anchor=east] {$U_1 = 0$};
  \node[left=1mm of m2-2-1.west, anchor=east] {$U_1 = 1$};
  \node[above=1mm of m2-1-1.north, anchor=south] {$U_2 = 0$};
  \node[above=1mm of m2-1-2.north, anchor=south] {$U_2 = 1$};
  \node[below=0.8mm of m2.south, anchor=north] {$X_0 = 1$};

  \matrix[matrixstyle, below=1.0cm of m1] (m3) {
      0 &0 \\
       \frac{1}{2} & 0 \\
  };
  \node[left=1mm of m3-1-1.west, anchor=east] {$U_1 = 0$};
  \node[left=1mm of m3-2-1.west, anchor=east] {$U_1 = 1$};
  \node[above=1mm of m3-1-1.north, anchor=south] {$U_2 = 0$};
  \node[above=1mm of m3-1-2.north, anchor=south] {$U_2 = 1$};
  \node[below=0.8mm of m3.south, anchor=north] {$X_0 = 1$};

  \matrix[matrixstyle, below=1.0cm of m2] (m4) {
      0 &\frac{1}{2} \\
       0 & 0 \\
  };
  \node[left=1mm of m4-1-1.west, anchor=east] {$U_1 = 0$};
  \node[left=1mm of m4-2-1.west, anchor=east] {$U_1 = 1$};
  \node[above=1mm of m4-1-1.north, anchor=south] {$U_2 = 0$};
  \node[above=1mm of m4-1-2.north, anchor=south] {$U_2 = 1$};
  \node[below=0.8mm of m4.south, anchor=north] {$X_0 = 1$};

  \matrix[matrixstyle, below=1.0cm of m3] (m5) {
      \frac{1}{4} &0 \\
       \frac{1}{4} & 0 \\
  };
  \node[left=1mm of m5-1-1.west, anchor=east] {$U_1 = 0$};
  \node[left=1mm of m5-2-1.west, anchor=east] {$U_1 = 1$};
  \node[above=1mm of m5-1-1.north, anchor=south] {$U_2 = 0$};
  \node[above=1mm of m5-1-2.north, anchor=south] {$U_2 = 1$};
  \node[below=0.8mm of m5.south, anchor=north] {$X_0 = 0$};

  \matrix[matrixstyle, below=1cm of m4] (m6) {
      0 &\frac{1}{4} \\
       0 & \frac{1}{4} \\
  };
  \node[left=1mm of m6-1-1.west, anchor=east] {$U_1 = 0$};
  \node[left=1mm of m6-2-1.west, anchor=east] {$U_1 = 1$};
  \node[above=1mm of m6-1-1.north, anchor=south] {$U_2 = 0$};
  \node[above=1mm of m6-1-2.north, anchor=south] {$U_2 = 1$};
  \node[below=0.8mm of m6.south, anchor=north] {$X_0 = 1$};

\end{tikzpicture}
\caption{First row: joint distribution given by $(\alpha = 0,\beta = 0)$, second row: joint distribution given by $(\alpha = 1,\beta = 1)$, third row: a convex combination of the above two cases, but is not achievable by any single pair of $(\alpha,\beta)$.}

%
  \label{fig: binary example, 0,1}
\end{figure}

    \end{example}

 \subsection{Causal-causal with Channel Feedback}

\begin{figure}[!t]
  \centering


\begin{tikzpicture}[scale=0.9, every node/.style={scale=0.9}]
    \draw (2,0) rectangle (3,1);
    \draw (6.8,0) rectangle (7.8,1);

    \draw (4.2,0.5) circle (0.2) node {$+$};
    \draw (5.6,0.5) circle (0.2) node {$+$};

    \filldraw (1,-0.5) circle (2pt) node[left] {$X_{0,t}$};
    \filldraw (5.6,1.5) circle (2pt) node[above] {$Z_{1,t}$};

    \draw[->] (1,-0.5) -- (1,0.5) -- (2,0.5);
    \draw[->] (1,-0.5) -- (4.2,-0.5) -- (4.2,0.3);
    \draw[->] (3,0.5) -- (4,0.5);
    \draw[->] (4.4,0.5) -- (5.4,0.5);
    \draw[->] (4.9,0.5) -- (4.9,-0.5) -- (8.8,-0.5);
    \draw[->] (5.6,1.5) -- (5.6,0.7);
    \draw[->] (5.8,0.5) -- (6.8,0.5);
    \draw[->] (7.8,0.5) -- (8.8,0.5);
    \draw[->] (6.3,1) -- (6.3, 2.1) -- (2.5,2.1) -- (2.5,1); 
     \draw[dotted, ->] (0.6, -0.2) -- (0.6, 2.3) -- (7.3,2.3) -- (7.3, 1); 

    \node at (1.5,0.8) {$X_0^t$};
    \node at (3.5,0.8) {$U_{1,t}$};
    \node at (4.9,0.8) {$X_{1,t}$};
    \node at (6.3,0.8) {$Y_{1}^t$};
    \node at (3,1.5) {$Y_{1}^{t-1}$};
    \node at (7.9,1.7) {$X_{0}^{t-1}$};
    \node at (8.3,0.8) {$U_{2,t}$};
    \node at (8.3,-0.2) {$X_{1,t}$};
    \node at (2.5,0.5) {$C_1$};
    \node at (7.3,0.5) {$C_2$};
\end{tikzpicture}



\caption{Witsenhausen counterexample for causal-encoding and causal-decoding with channel feedback $Y_1^{t-1}$ to the encoder. The dotted line describes the source feed-forward $X_0^{t-1}$ to the decoder, that is used for the genie-aided argument in the converse proof of Theorem \ref{theorem: c-c w-f} in Appendix B.}
\label{fig:c-c w-f model} 
  \label{fig: c-c w-f}
\end{figure}
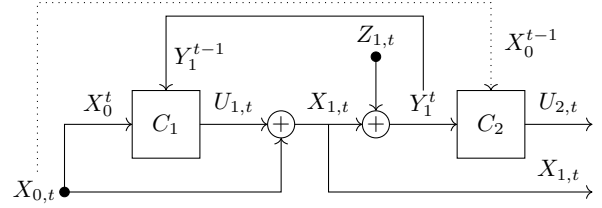

    Now suppose the causal encoder has a delayed access to the sequence of channel output $Y_1^{t-1}$ at each stage $t$ to help its decision, see Fig. \ref{fig: c-c w-f}.
    
\begin{definition}\label{def: c-c w-f}
    For $n\in\mathbb{N}$, a ``control design'' with causal encoder and causal decoder with channel feedback is a tuple of stochastic functions $c = (\{ f^{(\mathsf{f},t)}_{U_{1,t}|X_0^t,Y_1^{t-1}}\}_{t=1}^n, \{ g^{(t)}_{U_{2,t}|Y_1^t}\}_{t=1}^n)$ defined by
    \begin{equation}
        f^{(\mathsf{f},t)}_{U_{1,t}|X_0^t,Y_1^{t-1}}: \mathcal{X}_0^t\times \mathcal{Y}_1^{t-1} \longrightarrow \mathcal{U}_{1,t},\quad g^{(t)}_{U_{2,t}|Y_1^t}: \mathcal{Y}_1^t\longrightarrow \mathcal{U}_{2,t},
    \end{equation}
    which induces a distribution over sequences of symbols:
    \begin{equation}
         \prod_{t=1}^n \mathcal{P}_{X_{0,t}} \prod_{t=1}^n f^{(\mathsf{f},t)}_{U_{1,t}|X_0^t,Y_1^{t-1}}\prod_{t=1}^n \mathcal{P}_{X_{1,t},Y_{1,t}|X_{0,t},U_{1,t}} \prod_{t=1}^n g^{(t)}_{U_{2,t}|Y_1^t}, \label{eq: c-c w-f distribution of sequences}
    \end{equation}
where $Y_1^0 = \emptyset$. We denote by $\mathcal{C}_{ed,\mathsf{f}}(n)$ the set of control designs with causal encoder and causal decoder with channel feedback.
\end{definition}

We define the achievable pairs of cost $(P,S)$ similarly as in Definition \ref{def: achievable cost} and we denote by $\mathcal{R}_{ed, \mathsf{f}}$ the region of achievable pairs of costs.

\begin{theorem}\label{theorem: c-c w-f}
    $\mathcal{R}_{ed, \mathsf{f}} = \mathcal{R}_{ed}$.
\end{theorem}

The proof of  Theorem \ref{theorem: c-c w-f} is stated in Appendix B, where a genie-aided method is used to prove the converse result.

Adding channel feedback does not help for enlarging the region of achievable pairs of costs when both DMs are causal. This result is a consequence of the two causalities such that the two DMs only deal with new information: Even though at time $t\in[1:n]$ the encoder has the opportunity to provide more insights for its previous decisions based on the past channel output $Y^{t-1}_1$, the causal decoder has already responded to its past actions and therefore does not allow further modifications for this time. 

\subsection{Causal-noncausal with Channel Feedback}

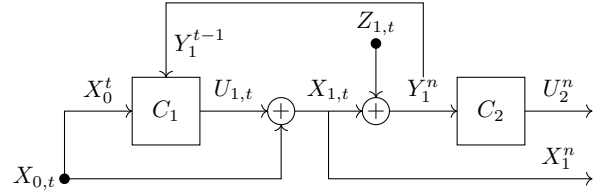
\begin{figure}[!t]
\vspace{0.5cm}
  \centering


\begin{tikzpicture}[scale=0.9, every node/.style={scale=0.9}]
    \vspace{1cm}
    \draw (2,0) rectangle (3,1);
    \draw (6.8,0) rectangle (7.8,1);

    \draw (4.2,0.5) circle (0.2) node {$+$};
    \draw (5.6,0.5) circle (0.2) node {$+$};

    \filldraw (1,-0.5) circle (2pt) node[left] {$X_{0,t}$};
    \filldraw (5.6,1.5) circle (2pt) node[above] {$Z_{1,t}$};

    \draw[->] (1,-0.5) -- (1,0.5) -- (2,0.5);
    \draw[->] (1,-0.5) -- (4.2,-0.5) -- (4.2,0.3);
    \draw[->] (3,0.5) -- (4,0.5);
    \draw[->] (4.4,0.5) -- (5.4,0.5);
    \draw[->] (4.9,0.5) -- (4.9,-0.5) -- (8.8,-0.5);
    \draw[->] (5.6,1.5) -- (5.6,0.7);
    \draw[->] (5.8,0.5) -- (6.8,0.5);
    \draw[->] (7.8,0.5) -- (8.8,0.5);
    \draw[->] (6.3,1) -- (6.3, 2.1) -- (2.5,2.1) -- (2.5,1); 

    \node at (1.5,0.8) {$X_0^t$};
    \node at (3.5,0.8) {$U_{1,t}$};
    \node at (4.9,0.8) {$X_{1,t}$};
    \node at (6.3,0.8) {$Y_{1}^n$};
    \node at (3,1.5) {$Y_{1}^{t-1}$};
    \node at (8.3,0.8) {$U_{2}^n$};
    \node at (8.3,-0.2) {$X_{1}^n$};
    \node at (2.5,0.5) {$C_1$};
    \node at (7.3,0.5) {$C_2$};
\end{tikzpicture}



\caption{Witsenhausen counterexample for causal-encoding and noncausal-decoding with channel feedback $Y_1^{t-1}$ to the encoder.}
\label{fig:c-n w-f model} 
  \label{fig: c-n w-f}
\end{figure}

In this section, we consider the model of causal-encoding and noncausal-decoding with channel feedback as illustrated in Fig. \ref{fig: c-n w-f}. The main result in this chapter provides the information constraint that completes the proof of Corollary III.6 in \cite{zhao2024CDC}.

\begin{definition}
    For $n\in\mathbb{N}$, a ``control design'' with causal encoder and noncausal decoder with channel feedback is a tuple of stochastic functions $c = (\{ f^{(\mathsf{f},t)}_{U_{1,t}|X_0^t,Y_1^{t-1}}\}_{t=1}^n, g_{U_{2}^n|Y_1^n})$ defined by
    \begin{equation}
        f^{(\mathsf{f},t)}_{U_{1,t}|X_0^t,Y_1^{t-1}}: \mathcal{X}_0^t\times \mathcal{Y}_1^{t-1} \longrightarrow \mathcal{U}_{1,t},\quad g_{U_{2}^n|Y_1^n}: \mathcal{Y}_1^n\longrightarrow \mathcal{U}_{2}^n,
    \end{equation}
    which induces a distribution over sequences of symbols:
    \begin{equation}
         \prod_{t=1}^n \mathcal{P}_{X_{0,t}} \prod_{t=1}^n f^{(\mathsf{f},t)}_{U_{1,t}|X_0^t,Y_1^{t-1}}\prod_{t=1}^n \mathcal{P}_{X_{1,t},Y_{1,t}|X_{0,t},U_{1,t}} g_{U_{2}^n|Y_1^n}, \label{eq: c-n w-f distribution of sequences}
    \end{equation}
where $Y_1^0 = \emptyset$. We denote by $\mathcal{C}_{e,\mathsf{f}}(n)$ the set of control designs with causal encoder and noncausal decoder with channel feedback.
\end{definition} 

Again, we denote by $\mathcal{R}_{e, \mathsf{f}}$ the region of achievable pairs of rates for control designs in $\mathcal{C}_{e,\mathsf{f}}(n)$.

\begin{theorem}\label{theorem: c-n w-f}
    The pair of Witsenhausen costs is achievable $(P,S)$ if and only if there exists a joint distribution over the random variables $(X_0, W_1, U_1, X_1, Y_1, U_2)$ that decomposes according to
    \begin{equation}
\mathcal{P}_{X_0}\mathcal{P}_{W_1}\mathcal{P}_{U_1|X_0,W_1}\mathcal{P}_{X_1, Y_1|X_0, U_1}\mathcal{P}_{U_2|X_0, W_1, Y_1},\label{eq: c-n w-f prob result}
    \end{equation}
    such that
    \begin{align}
        &I(W_1; Y_1) - I(U_2; X_0 | W_1,Y_1) \geq 0,\label{eq: c-n w-f info result}\\
        &P = \mathbb{E}\sbrackets{U_1^2}, \quad \quad S = \mathbb{E}\sbrackets{(X_1 - U_2)^2},\label{eq: c-n w-f cost result}
    \end{align}
    where $\mathcal{P}_{X_0}$ and $\mathcal{P}_{X_1, Y_1|X_0, U_1}$ are given Gaussian distributions, and $W_1$ is an auxiliary random variables.
\end{theorem}

\begin{remark}
Distribution \eqref{eq: c-n w-f prob result} satisfies the follows
    \begin{align}
        \left\{  
        \begin{aligned}
            &X_0\text{ is independent of }W_1,\\
            &(X_1,Y_1)-\!\!\!\!\minuso\!\!\!\!- (X_0, U_1)    -\!\!\!\!\minuso\!\!\!\!- W_1,\\
            & U_2 -\!\!\!\!\minuso\!\!\!\!- (X_0, Y_1,W_1) -\!\!\!\!\minuso\!\!\!\!- (U_1, X_1).
        \end{aligned}
        \right.
        \label{c-n w-f markov result}
        \end{align}
\end{remark}

\begin{remark}
    Compared to the single-letter result in \cite[Theorem II.3]{zhao2024coordination}, the presence of channel feedback enables the decoder to directly coordinate with the source state $X_0$ instead of its noisy representation ($W_2$ in the reference paper). Therefore, feedback enlarges the set of achievable pairs. A similar observation has been also pointed out in \cite{bross2017rate, Letreust2015empirical, Letreust2021state}
\end{remark}

\begin{proof}[Sketch of the proof for Theorem \ref{theorem: c-n w-f}]

The converse proof for Theorem \ref{theorem: c-n w-f} can be adapted from the converse proof of Theorem III.2 in \cite{Letreust2015empirical}, which dealt with a state-independent channel. We can modify the arguments from this reference to apply to the state-dependent channel in our setting, without affecting the outcome of the result.


Additionally, the coding scheme for the achievability proof for Theorem \ref{theorem: c-n w-f} also extends the approach provided in \cite{Letreust2015empirical} with changing the channel from state-independent to state-dependent. The cost analysis is directly derived from the arguments presented in \cite{zhao2024coordination}.   
\end{proof}

\section*{Appendix A: Proof of Theorem \ref{theorem: c-c wits main}}
\begin{proof}[Converse]
For a pair $(P,S)$, assume that we have a control design $c\in\mathcal{C}_{ed}(n)$ of block length $n\in\mathbb N$ and small $\varepsilon>0$ which induces a joint p.m.f. $\mathcal{P}^n$ of the form \eqref{eq: distribution of sequences} such that
    \begin{align}
        \abs{P - \mathbb E[c_P(U_1^n)]
        } + \abs{S - \mathbb E[c_S(X_1^n,U_2^n)]}<\varepsilon. \label{eq: def of achievable cost}
    \end{align}
    This is implied by condition \eqref{eq: achievable def}.

let $Q$ be an independent time random variable uniformly distributed over $\{1,...,n\}$. Define auxiliary random variables $X_0 = X_{0,Q}, U_1 = U_{1,Q}, X_1 = X_{1,Q}, Y_1 = Y_{1,Q}, U_2=U_{2,Q}$ with distribution
\begin{flalign}
&\mathbb{P}((X_0,U_1,X_1,Y_1,U_2) = (x_0,u_1,x_1,y_1,u_2)) \label{eq: distr aux RV}\\
&= \frac1n \sum_{q=1}^n\mathbb{P}((X_{0,q},U_{1,q},X_{1,q},Y_{1,q},U_{2,q}) = (x_0,u_1,x_1,y_1,u_2))\nonumber\\
&\quad\quad\quad\quad\quad\forall (x_0,u_1,x_1,y_1,u_2)\in\mathcal{X}_0\times\mathcal{U}_1\times\mathcal{X}_1\times\mathcal{Y}_1\times\mathcal{U}_2.\nonumber
\end{flalign}   
Then, the expected long-run costs could be reformulated as
\begin{flalign}
        \mathbb E[c_P(U_1^n)] &= \mathbb E\sbrackets{\frac{1}{n}\sum_{q=1}^nU_{1,q}^2} =\mathbb E[U_{1}^2], \label{eq: P reformulate}  \\  
     \mathbb E[c_S(X_1^n,U_2^n)]&
     = \mathbb E\sbrackets{\frac{1}{n}\sum_{q=1}^n(X_{1,q}-U_{2,q})^2} \\
     &= \mathbb E[(X_1 - U_2)^2]. \label{eq: S reformulate}  
\end{flalign}
Therefore, given \eqref{eq: def of achievable cost}, we obtain that
    \begin{align}
        |P - \mathbb E[U_{1}^2]| + |S - \mathbb E[(X_1 - U_2)^2]|<\varepsilon,
    \end{align}
    which is valid for all $\varepsilon\geq 0$. Hence, we have \eqref{eq: c-c cost result}. 

Now, we define the new auxiliary random variables $W_q = Y^{q-1}_1$ for $q=1,...,n$ and $T = (W_Q,Q)$. These auxiliary random variables satisfy the following Markov chains
\begin{itemize}
    \item $X_0\indep T$: This is because the source is i.i.d. generated, independent of the time stage $Q$ as well as the past channel output sequence $Y_1^{Q-1}$ due to causal encoding.
    \item $(X_1, Y_1)-\!\!\!\!\minuso\!\!\!\!- (X_0, U_1)    -\!\!\!\!\minuso\!\!\!\!- T$: This comes from the discrete memoryless channel, the characterization of the output $(X_1,Y_1)$ depends only on the input $(X_0,U_1)$.
    \item $U_2 -\!\!\!\!\minuso\!\!\!\!- (T, Y_1) -\!\!\!\!\minuso\!\!\!\!- (X_0, U_1, X_1)$: This is the consequence of the causal decoding that the reconstruction is fully characterized by sequence $Y_1^{Q}$ up to stage $Q=1,...,n$.
\end{itemize}

Therefore, the distribution of all the introduced auxiliary random variables decomposes as \eqref{c-c prob result}.
\end{proof}

\begin{proof}[Achievability]
Consider a joint distribution $\mathcal{P}$ of the form of \eqref{c-c prob result} with $\mathbb E[U_1^2] = P$ and $\mathbb E[(X_1 - U_2)^2] = S$. Fix a small $\varepsilon>0$ and a blocklength $n\in\mathbb N$.

The encoder and decoder simply conduct symbol-by-symbol approaches based on the given distribution $\mathcal{P}$: Before the transmission, the encoder selects a typical sequence $T^n\in\mathcal{T}_\varepsilon(\mathcal{P}_T)$ and shares it to the decoder. Then, at each time $t\in\{1,...,n\}$, the encoder observes $X_{0,t}$, and outputs $U_{1,t}\sim\mathcal{P}_{U_1|X_0,T}(\cdot|X_{0,t},T_t)$. The channel generates $X_{1,t},Y_{1,t}\sim\mathcal{P}_{X_1,Y_1|X_0,U_1}(\cdot|X_{0,t},U_{1,t})$. Then, the decoder draws $U_{2,t}\sim\mathcal{P}_{U_2|T,Y_1}(\cdot|T_t,Y_{1,t})$.

In such case, since each symbol is generated i.i.d. according to its distribution, from the law of large numbers (LLN), we have 
\begin{align}
    &c_P(U_1^n) = \frac{1}{n}\sum_{t=1}^nU_{1,t}^2\xrightarrow[]{n\rightarrow\infty}  P.\\
    &c_S(X_1^n, U_2^n) = \frac{1}{n}\sum_{t=1}^n(X_{1,t} - U_{2,t})^2\xrightarrow[]{n\rightarrow\infty} S
\end{align}

Since convergence in probability implies convergence in $\mathscr{L}^1$ measure, given the existence of finite second moment, we have
\begin{align}
    &\mathbb{E}[|c_P(U_1^n) - P|]<\frac{1}{2}\varepsilon,\\
    & \mathbb{E}[|c_S(X_1^n, U_2^n) - S|]<\frac{1}{2}\varepsilon.
\end{align}
for sufficiently large $n$. 
\end{proof}

\begin{proof}[Proof of the Cardinality Bound of $T$]
Consider the set $\mathscr{P}$ of all p.m.f.s on $\mathcal{U}_1\times\mathcal{X}_1\times\mathcal{U}_2$ satisfying the form of \eqref{c-c prob result} and the following two continuous functions
    \begin{align}
        f_1(\mathcal{P},t) &= \mathbb E\sbrackets{U_1^2\mid T=t},\\
        f_2(\mathcal{P},t) &= \mathbb E\sbrackets{(X_1-U_2)^2\mid T=t}.
    \end{align}
    From the support lemma in \cite{csiszar2011information}, \cite[Appendix C]{elgamal2011nit}, we can establish the time-sharing cardinality bound $|\mathcal{T}|\leq 2$ for the convex-hull operation.
    
\end{proof}

\section*{Appendix B: Proof of Theorem \ref{theorem: c-c w-f}}
\begin{proof}[Converse]

Consider a control design illustrated in Fig. \ref{fig: c-c w-f} together with the dotted line. This
induces a distribution of the following form
\small\begin{align}
     \prod_{t=1}^n \mathcal{P}_{X_{0,t}} \prod_{t=1}^n f^{(\mathsf{f},t)}_{U_{1,t}|X_0^t,Y_1^{t-1}}\prod_{t=1}^n \mathcal{P}_{X_{1,t},Y_{1,t}|X_{0,t},U_{1,t}} \prod_{t=1}^n g^{(\mathsf{f},t)}_{U_{2,t}|Y_1^t,X_0^{t-1}}. \label{eq: c-c w-f-f distribution of sequences}
\end{align}
\normalsize
This control design is clearly more powerful than that in Definition \ref{def: c-c w-f} because the causal decoder also receives a past sequence of source information $X_0^{t-1}$ at each instant $t$. Therefore, the achievable cost region induced by this superior system of adding the source feed-forward serves as an outerbound for the region without the source feed-forward. We denote this new region by $\mathcal{R}_{ed, \mathsf{f},\mathsf{f}}$ and we have
\begin{align}\label{eq: cost region subsets}
    \mathcal{R}_{ed, \mathsf{f},\mathsf{f}}\supseteq \mathcal{R}_{ed, \mathsf{f}}\supseteq\mathcal{R}_{ed}
\end{align}

Now, we show that $\mathcal{R}_{ed, \mathsf{f},\mathsf{f}}= \mathcal{R}_{ed}$.

Similar to the converse proof of Theorem \ref{theorem: c-c wits main} in Appendix A, we introduce a time-sharing random variable $Q\sim\text{Unif}[1,...,n]$ and define a sequence of new random variables $X_0 = X_{0,Q}, U_1 = U_{1,Q}, X_1 = X_{1,Q}, Y_1 = Y_{1,Q}, U_2 = U_{2,Q}$ with their joint distribution \eqref{eq: distr aux RV}. In this way, the $n$-stage long-run costs for the control design of \eqref{eq: c-c w-f-f distribution of sequences} also could be reformulated to \eqref{eq: P reformulate} and \eqref{eq: S reformulate}.

Next, let $W_q = (X_0^{q-1}, Y_1^{q-1})$ for $q=1,...,n$ and $T = (W_Q, Q)$ be new auxiliary random variables. Then, it holds that
\begin{itemize}
    \item $X_0\indep T$. This follows from the i.i.d. source and causal encoding.
    \item $(X_1, Y_1)-\!\!\!\!\minuso\!\!\!\!- (X_0, U_1)    -\!\!\!\!\minuso\!\!\!\!- T$: This comes from the discrete memoryless channel.
    \item $U_2 -\!\!\!\!\minuso\!\!\!\!- (T, Y_1) -\!\!\!\!\minuso\!\!\!\!- (X_0, U_1, X_1)$: This is the consequence of the causal decoding at the presence of $Y_1^{Q-1}, X_0^{Q-1}$ at each time instant $Q$.
\end{itemize}

Given the above Markov chains, the single-letter joint distribution that characterizes a desired control scheme of adding a source feed-forward also decomposes as \eqref{c-c prob result}. 

Therefore, we have shown that $\mathcal{R}_{ed, \mathsf{f},\mathsf{f}}= \mathcal{R}_{ed}$. Namely, adding both the channel feedback AND the source feed-forward information does not enlarge the achievable Witsenhausen cost region of $\mathcal{R}_{ed}$. Since we also have \eqref{eq: cost region subsets}, by the genie-aided argument, we conclude $\mathcal{R}_{ed, \mathsf{f}} = \mathcal{R}_{ed}$.\end{proof}


\bibliographystyle{ieeetr}
\bibliography{main}

\end{document}